\documentclass[debug]{rmaa}

\usepackage{paralist}
\usepackage[normalem]{ulem}
\usepackage{psfrag,color}
\usepackage[dvipsnames]{xcolor}
\usepackage{hyperref} 
\hypersetup{
  colorlinks = true, 
  urlcolor   = blue, 
  linkcolor  = blue, 
  citecolor  = blue  
}
\usepackage[latin1]{inputenc}

\newcommand{\zyla}{Andor Zyla 4.2-Plus USB 3.0}

\newcommand{\add}[1]{{\color{blue}{\textbf\small{#1}}}}

\title{OPTICAM: A triple-camera optical system designed to explore the fastest timescales in Astronomy} 

\author{
  A.~Castro,\altaffilmark{1,2,$\dagger$} 
  D.~Altamirano,\altaffilmark{1} 
  R.~Michel,\altaffilmark{2} 
  P.~Gandhi,\altaffilmark{1} 
  J.~V.~Hern{\'a}ndez~Santisteban,\altaffilmark{1,4,5} 
  J.~Echevarr{\'i}a,\altaffilmark{3} 
  C.~Tejada,\altaffilmark{2} 
  C.~Knigge,\altaffilmark{1} 
  G.~Sierra,\altaffilmark{2} 
  E.~Colorado,\altaffilmark{2} 
  J.~Hern{\'a}ndez-Landa,\altaffilmark{2} 
  D.~Whiter,\altaffilmark{1} 
  M.~Middleton,\altaffilmark{1} 
  B.~Garc{\'i}a,\altaffilmark{2},
  G.~Guisa,\altaffilmark{2}
  and N.~Castro~Segura\altaffilmark{1}
}

\altaffiltext{1}{Physics and Astronomy, University of Southampton, Southampton, UK}
\altaffiltext{2}{Instituto de Astronom{\'i}a, UNAM, Ensenada, M{\'e}xico}
\altaffiltext{3}{Instituto de Astronom{\'i}a, UNAM, Ciudad de M{\'e}xico, M{\'e}xico}
\altaffiltext{4}{Anton Pannekoek Institute, University of Amsterdam, Amsterdam, the Netherlands}
\altaffiltext{5}{SUPA School of Physics \& Astronomy, University of St. Andrews, St. Andrews, UK}
\altaffiltext{6}{$\dagger$Royal Society Newton International Fellow}
\shortauthor{Castro et al.}
\shorttitle{OPTICAM: Simultaneous triple-camera imaging system.}
\fulladdresses{
\item Angel Castro, Diego Altamirano, Poshak Gandhi, Christian Knigge, Daniel Whiter, Matthew J. Middleton and Noel Castro-Segura: Physics and Astronomy, University of Southampton, SO17 1BJ, Southampton, UK (a.castro, d.altamirano, poshak.gandhi, c.knigge, d.whiter, m.j.middleton, n.castro-segura@soton.ac.uk).

\item Ra{\'u}l Michel, Carlos Tejada, Gerardo Sierra, Enrique Colorado, Javier Hern{\'a}ndez Landa, Gerardo Guisa and Benjam{\'i}n Garc{\'i}a: Instituto de Astronom{\'i}a, Universidad Nacional Aut{\'o}noma de M{\'e}xico, 22860, Ensenada, M{\'e}xico (rmm, ctejada, gerardo, landa, guisa, benji@astro.unam.mx).

\item Juan Echevarr{\'i}a: Instituto de Astronom{\'i}a, Universidad Nacional Aut{\'o}noma de M{\'e}xico, Ciudad de M{\'e}xico, M{\'e}xico (jer@astro.unam.mx).

\item Juan~V.~Hern{\'a}ndez Santisteban: SUPA School of Physics \& Astronomy, University of St. Andrews, North Haugh, St. Andrews KY16 9SS, UK (jvhs1@st-andrews.ac.uk).

}

\listofauthors{A. Castro et al.}

\indexauthor{Castro, A.}
\indexauthor{Altamirano, D.}
\indexauthor{Michel, R.}
\indexauthor{Gandhi, P.}
\indexauthor{Hern{\'a}ndez Santisteban,~J.V.}
\indexauthor{Echevarr{\'i}a, J.}
\indexauthor{Tejada, C.}
\indexauthor{Knigge, C.}
\indexauthor{Sierra, G.}
\indexauthor{Colorado, E.}
\indexauthor{Hern{\'a}ndez-Landa, J.}
\indexauthor{Whiter, D.}
\indexauthor{Middleton, M.}
\indexauthor{Garc{\'i}a, B.}
\indexauthor{Guisa, G.}
\indexauthor{Castro-Segura, N.}

\abstract{We report the development of a high-time resolution, 3-colour, simultaneous optical imaging system for the 2.1 m telescope in San Pedro M{\'a}rtir Observatory, M{\'e}xico. OPTICAM will be equipped with three Andor Zyla 4.2-Plus sCMOS cameras and a set of SDSS filters allowing optical coverage in the 320--1,100 $\mathrm{nm}$ range. OPTICAM will nominally allow sub-second exposures. Given its instrumental design, a wide range of fast-variability astrophysical sources can be targeted with OPTICAM including X-ray binaries, pulsating white dwarfs, accreting compact objects, eclipsing binaries and exoplanets. OPTICAM observations will be proprietary for only six months and will then be made publicly available for the astronomical community.
}

\resumen{
Informamos sobre el desarrollo de un sistema de imagen {\'o}ptica simult{\'a}neo en 3 colores y alta resoluci{\'o}n temporal para el telescopio de 2.1 m del Observatorio Astron{\'o}mico Nacional en San Pedro M{\'a}rtir, M{\'e}xico. OPTICAM estar{\'a} equipado con tres c{\'a}maras sCMOS Andor Zyla 4.2-Plus y un conjunto de filtros SDSS que permitir{\'a}n cobertura {\'o}ptica en el rango de 320--1,100 $\mathrm{nm}$. OPTICAM permitir{\'a} nominalmente exposiciones en el orden de sub-segundos. Dado su dise\~no instrumental, un amplio rango de fuentes astrof{\'i}sicas de r{\'a}pida variabilidad pueden ser observadas con OPTICAM incluyendo estrellas binarias de rayos X, enanas blancas pulsantes, objetos compactos en acrecentamiento, estrellas binarias eclipsantes y expolanetas. Las observaciones realizadas con OPTICAM ser{\'a}n de propiedad exclusiva durante seis meses y luego se pondr{\'a}n a disposici{\'o}n p{\'u}blica para la comunidad astron{\'o}mica.

}
\addkeyword{instrumentation: photometers}
\addkeyword{stars: imaging}
\addkeyword{telescopes}
\addkeyword{accretion, accretion discs}

\begin{document}
\maketitle


\section{General}
\label{sec:intro}

Many physical processes occurring on short time scales (on the order of milliseconds to seconds) have only recently been observed thanks to the advent of new technologies, allowing multi-wavelength high-cadence photometric measurements. As a consequence of these advances, along with the development of new observational techniques, time-domain astronomy has become a strong emerging area in modern astrophysics.

High time-resolution observations have revolutionised our view of the physical processes that take place on short timescales, providing a unique insight into physical phenomena that otherwise would be inaccessible to conventional photometry. For example, the vast majority of all known extra-solar planets have been discovered by searching for transits in the optical light curves of their host stars \citep[e.g.][]{mayor+95,pepe+14}; neutron star (NS) spin periods ranging from milli-seconds to hours have been discovered by observing their X-ray and optical light curves \citep[e.g.][]{ambrosino+17}; accreting astrophysical systems from young stars and white dwarfs (WDs), to stellar black holes (BHs) in binary systems and supermassive black holes (SMBHs) in quasars have been found to display the same stochastic 'flickering' variability \citep{mchardy+06,scaringi+15}; and evidence for strong-field general relativistic precession has been found in accreting stellar-mass BHs via the detection of fast, quasi-periodic oscillations \citep[QPOs;][]{motta+14}. 

Accretion-induced variability manifests through changes in the brightness of an astrophysical object on different time-scales and across a range of wavelengths. Recent research by \citet{scaringi+15} suggests that the characteristic time-scale of these variations depends mainly on the physical properties of the accretor and the rate at which it accretes, regardless of the nature of the accretor. Accreting WDs, NSs and BHs in binary systems provide us with the best available laboratories for studying the process of disc accretion. Particularly in the innermost disc regions, the variability time-scales can range from seconds in WDs to milli-seconds in NSs and stellar-mass BHs \citep{dhillon+07}. Multi-wavelength high-speed observations are essential to increase our understanding of the underlying physics of the accretion process \citep[see][for review]{middleton+17}. Unfortunately, in-depth and systematic studies in the optical wavelength range have not been possible due to the lack of instruments with the necessary temporal resolution and sensitivity. 

Over the past decade, fast-photometric studies have provided a glimpse of their potential when used in simultaneous multi-wavelength campaigns \citep[e.g.][]{gandhi+10,scaringi+13,shahbaz+13,mallonn+15,gandhi+16,gandhi+17,hynes+19,malzac+18,mouchet+17,green+18,paice+19,pala+19}. For example, in accreting compact objects in X-ray binaries (XRBs), the relationship between the rapid (millisecond to minutes) variability at different wavelengths has allowed the inner structure close to the event horizon of a BH or the surface of a NS to be determined through measurements of the absolute sizes of its main components and through inferences about the plasma physics underlying the observed phenomena \citep[e.g.][]{gandhi+08,gandhi+16}.

The relationship between optical and X-ray variability is known to change as a function of time and/or spectral state in both BH and NS XRBs, suggesting changes in the physical scale and geometry of the emitting regions \citep[][and references therein]{kanbach+01,durant+08, gandhi+08, gandhi+10, gandhi+17}. As a consequence of the lack of current facilities providing fast-photometric, multi-band capabilities, there are only a few simultaneous observations of such systems, and even fewer observations made with high temporal resolution. In this context, fundamental contributions have been recently made through the use of the ULTRACAM \citep{dhillon+07} instrument in several key areas, ranging from accretion physics to exoplanets. Major contributions are also expected from its successor HiPERCAM \citep{dhillon+16,dhillon+18,bezawada+18}, a state-of-the-art quintuple-beam imager for high time-resolution observations. Other short-term events such as the transit of exoplanets, have been targeted by MuSCAT2 \citep{narita+19} using simultaneous 4-colour imaging at the 1.52 m Carlos S{\'a}nchez telescope. The TAOS project \citep[e.g.][]{bianco+10,geary+12,lehner+12,ricci+14} has focused on the comprehensive study of Kuiper Belt Objects (KBOs) through measurements in the sub-second range. KBOs have also been successfully targeted by CHIMERA \citep{harding+16}, a wide-field, simultaneous, two-color, high-speed photometer on the Hale telescope.

As part of this new generation of time-domain instruments, we present OPtical TIming CAMera (OPTICAM\footnote{\url{https://www.southampton.ac.uk/opticam}}), a new high-speed, triple-beam, optical imaging system developed through a collaboration between research teams at the Universidad Nacional Aut{\'o}noma de M{\'e}xico (UNAM) and the Astronomy Group at the University of Southampton (AG-UoS) in the United Kingdom. OPTICAM will open new areas of opportunity for both research communities which will benefit by accessing an instrument equipped with three high-sensitivity, high-resolution, state-of-the-art scientific complementary metal-oxide-semiconductor (sCMOS) cameras. OPTICAM will be equipped with a set of Sloan Digital Sky Survey (SDSS) filters and three $2\,048\times2\,048$ sCMOS detectors. The fastest timescale variations likely to be observed with OPTICAM range from milliseconds to seconds depending on the apparent magnitude of the source. Simultaneous triple-band observations will allow problems due to stochastic variations to be avoided, while providing a broad wavelength coverage across the optical spectrum.
The instrument will be available as a common-user instrument for the 2.1 m telescope in the Mexican National Astronomical Observatory at San Pedro M{\'a}rtir, M{\'e}xico (OAN-SPM\footnote{\url{http://www.astrossp.unam.mx/oanspm/}}), which is among the best astronomical sites in the Northern Hemisphere (more details in \S~\ref{sec:spm_2m}). OPTICAM will be available to the Mexican and AG-UoS astronomical communities through traditional telescope time request methods. 

In this paper, we report the design and performance of OPTICAM, a triple-band high-speed optical system designed to explore sub-second variability scales. In \S~\ref{sec:spm_2m} we describe some of the characteristics of the OAN-SPM and the 2.1 m telescope. In \S~\ref{sec:opto-mechanical_design} a brief description of the opto-mechanical design is presented. Details about the utilised filter set and sCMOS camera are shown in \S~\ref{sec:filter_efficiency} and \S~\ref{sec:camera}, respectively. The flow-control and frame-acquisition system is described in \S~\ref{sec:control_data}. An estimate of the limiting magnitude for the SDSS optical bands is provided in Section \ref{sec:limiting_magnitude}. In \S~\ref{sec:science_opticam}, we explore some of the possible scientific cases that can be addressed using the OPTICAM instrument. Finally, in \S~\ref{sec:conclusions}, we present our conclusions and report on the current status and future time-line of OPTICAM.

\section{The OAN-SPM and the 2.1 m telescope}\label{sec:spm_2m}

The OAN-SPM is located on the Sierra San Pedro M{\'a}rtir in Ensenada, Baja California, M{\'e}xico (2800 m; 115\textdegree 27\arcmin 49\farcs ~W, 31\arcdeg 02\arcmin 39\farcs~N). The average annual temperature in SPM is 3 {\textdegree}C, with extremes between -10 {\textdegree}C and 20 {\textdegree}C and an average pressure of 0.74 atm. The OAN-SPM is one of the best observing sites in the Northern Hemisphere with 82\% clear sky nights \citep{carrasco+12}. Based on results from a 3-year study, \citet{echevarria+98} reported a median seeing of 0.61 arcsec and a first quartile of 0.50 arcsec. Later, \citet{michel+03} also reported a median seeing of 0.6 arcsec using the Differential Image Motion Monitor (DIMM) technique. Additionally, \citet{tapia+03} reported that 63\% of the nights between 1984 and 2003 were considered ``photometric" whereas 73\% of nights were suitable for high-quality spectroscopic studies. 

An average seeing of 0.7 arcsec is estimated based on several long- and mid-term monitoring studies involving a variety of measurement techniques that took place at the OAN-SPM in recent years (see Table \ref{tab:spm_seeing}). These results also show a very stable, long-term sky quality at the site. In this sense, important efforts\footnote{\url{http://leydelcielo.astrosen.unam.mx/}} are being made with the intention of issuing regional legislation for the protection and conservation of the dark skies of northwestern M{\'e}xico. All of the aforementioned features make the OAN-SPM one of the best locations for optical Astronomy, comparable to other sites such as Roque de los Muchachos \citep[e.g.][]{munoz+97}, Las Campanas \citep{prieto+10}, Armazones and Mauna Kea \citep[e.g.][]{schock+09}.

The OAN-SPM 2.1 m telescope\footnote{\url{http://www.astrossp.unam.mx/~sectec/web/telescopios/2mt.html}} is the largest optical telescope in M{\'e}xico and started operations in 1979. The main mirror is fixed in an equatorial Ritchey-Chr{\'e}tien focus. The telescope has hyperbolic primary and secondary mirrors. Three focal lengths are available under request in the 2.1 m telescope: $f$/7.5 ($\sim$ 13.0 \arcsec/mm), $f$/13.5 ($\sim$ 7.15 \arcsec/mm) and $f$/30 ($\sim$ 3.25 \arcsec/mm). Recently, a fully automated instrument rotator has been permanently mounted on the telescope focus. OPTICAM will be installed on the Cassegrain focus of the 2.1 m telescope (see Fig.~\ref{fig:spm_2m_optical_path}) as a common-user instrument under the normal rules of operation and management currently in use at the OAN-SPM. 

\begin{table*}[!t]\centering
\footnotesize
  \setlength{\tabnotewidth}{0.6\columnwidth}
  \tablecols{5}
  \setlength{\tabcolsep}{1.0\tabcolsep}
  \caption{OAN-SPM seeing monitoring studies.} \label{tab:spm_seeing}
 \begin{tabular}{lcccc}
    \toprule
    Method\tabnotemark{a} & Reference & First Quartile & Median & Third Quartile \\
           &   & [arcsec]       & [arcsec] & [arcsec]  \\
    \midrule
    STT+CM   & \citet{echevarria+98} & 0.50  & 0.61 & \nodata\\
    DIMM      & \citet{conan+02}      & 0.61  & 0.77 & 0.99 \\
    DIMM      & \citet{michel+03}     & 0.48  & 0.60 & 0.81 \\
    MASS-DIMM & \citet{skidmore+09}   & 0.61  & 0.79 & 1.12 \\
    g-SciDAR  & \citet{avila+11}      & 0.50  & 0.68 & 0.97 \\
    DIMM      & \citet{sanchez+12}    & 0.60  & 0.78 & 1.11 \\
    MASS      & \citet{sanchez+12}    & 0.25  & 0.37 & 0.56 \\
    GL        & \citet{sanchez+12}    & 0.45  & 0.59 & 0.84 \\
    \bottomrule
    \tabnotetext{a}{Methods: Site Testing Telescope + Carnegie Monitor (STT+CM); Differential Image Motion Monitor (DIMM); Multi-Aperture Scintillation Sensor (MASS); Generalised-Scintillation Detection and Ranging (g-SciDAR); Ground Layer (GL).}
  \end{tabular}
\end{table*}

\section{Opto-mechanical design}
\label{sec:opto-mechanical_design}
\begin{figure*}[!t]
	\centering
  \includegraphics[width=0.6\columnwidth]{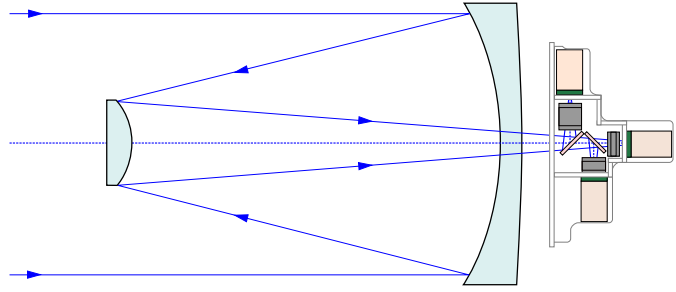}
  \caption{Layout of the optomechanical system of the OAN-SPM 2.1 m telescope and OPTICAM. Incident light is separated into three channels by two dichroic beam splitters. See also Fig.~\ref{fig:optical_design} and Fig.~\ref{fig:mechanical_config} for a more detailed description). The diagram is not to-scale but for illustrative purposes. 
  }
  \label{fig:spm_2m_optical_path}
\end{figure*}

\begin{figure*}[!t]
	\centering
  \includegraphics[width=0.5\columnwidth]{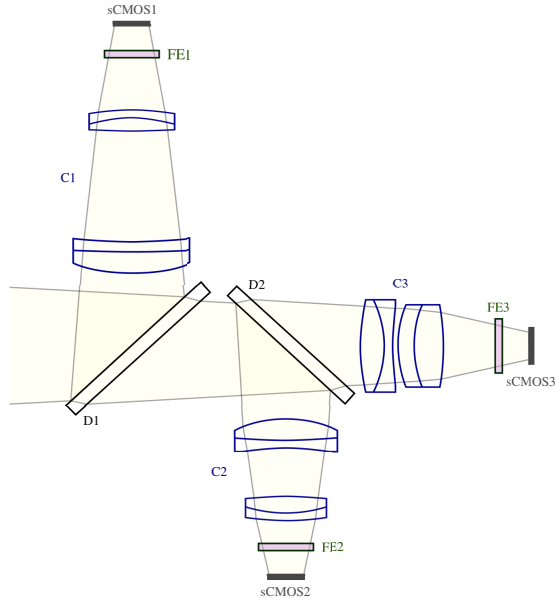}
  \caption{Design of the OPTICAM optical layout showing the major optical components: the dichroics (D1 and D2), optical cameras (C1, C2 and C3), filter exchangers (FE1 and FE2) and sCMOS cameras (sCMOS1, sCMOS2 and sCMOS3). This diagram is not to-scale.}
  \label{fig:optical_design}
\end{figure*}

OPTICAM is designed to operate at $f$/7.5 on the Cassegrain focus of the OAN-SPM 2.1 m telescope. The layout of the optical system of the 2.1 m telescope and OPTICAM is shown in Fig.~\ref{fig:spm_2m_optical_path}. OPTICAM will have three arms. The geometric layout of the optical design of OPTICAM is shown in Fig.~\ref{fig:optical_design}. The beam enters from the OAN-SPM 2.1 m telescope (from the left as shown in the Fig.~\ref{fig:optical_design} optical layout), passes through the first dichroic in Arm 1 where the initial beam is split. The reflected beam passes to the the first optical camera (C1). This camera is optimised for the use of $u'$ and $r'$ filters (from 320 to 550 $\mathrm{nm}$). In the same way, the transmitted beam is then split again towards the optical cameras C2 and C3, which are optimised for $r'$ (from 560 to 690 $\mathrm{nm}$), and $i'$ and $z'$ filters (from 690 to 1,100 $\mathrm{nm}$) in Arms 2 and 3, respectively. 

\begin{figure*}[!t]
	\centering
	\includegraphics[width=0.7\columnwidth]{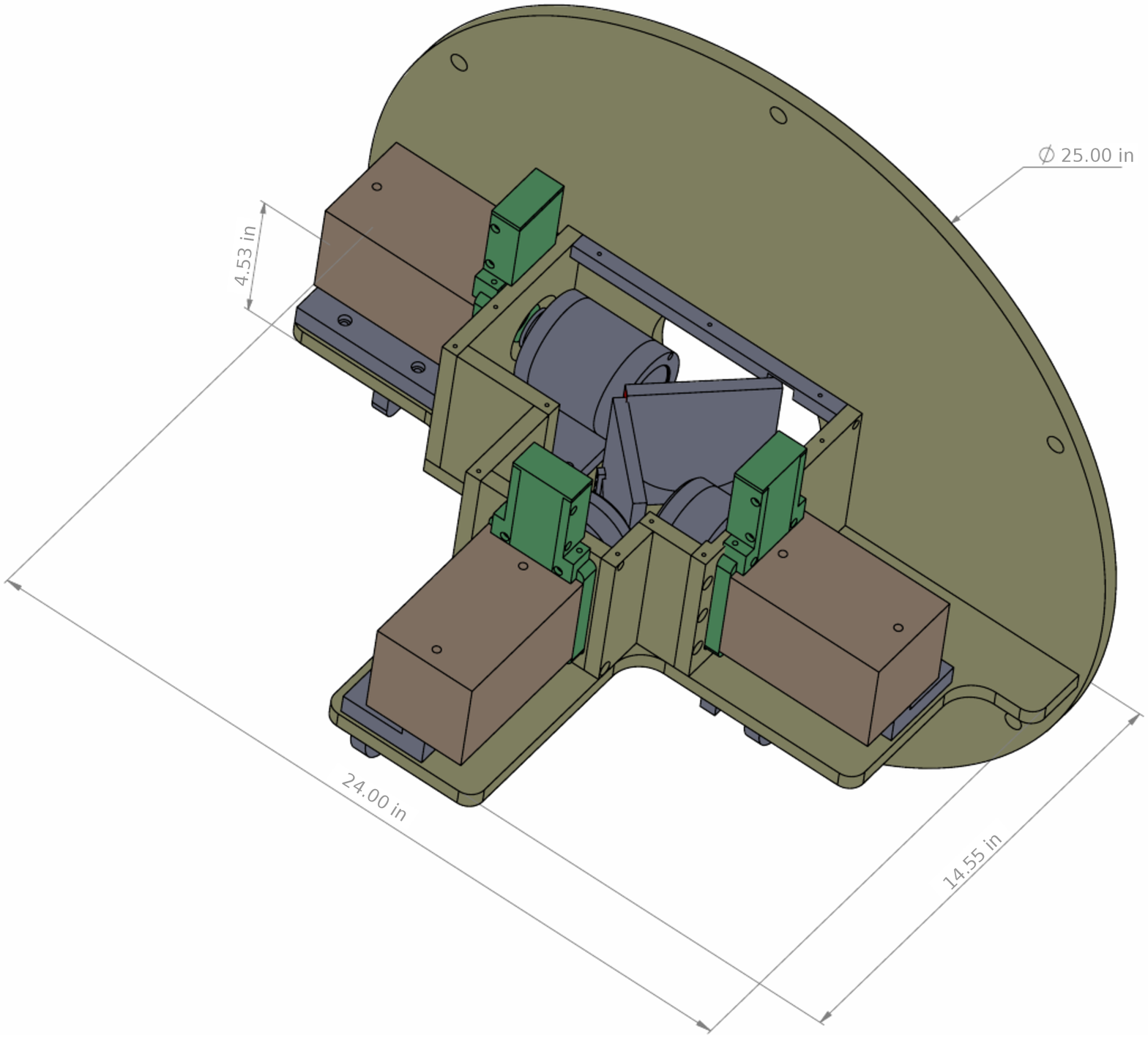}
	\caption{3D CAD  {\sc Solidworks} render of the OPTICAM instrumental setup. In this perspective, the top cover has been removed to show an internal view of the optical design. The dichroic beam splitter and optical lens barrels can be observed. Dimensions are given in inches.}
	\label{fig:mechanical_solidworks}
\end{figure*}

\begin{figure*}[!t]
	\centering
	\includegraphics[width=0.4\columnwidth]{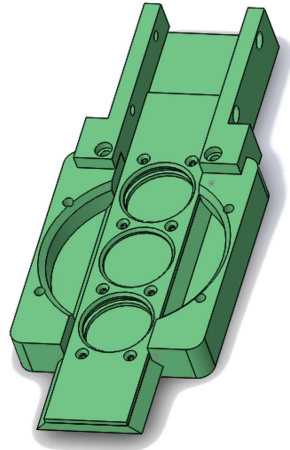}
	\caption{A manual filter exchanger will be used for each camera during the first stage of operations of OPTICAM.}
	\label{fig:filters_exchanger}
\end{figure*}

Custom filter exchangers (see Fig. \ref{fig:filters_exchanger} and \S~\ref{sec:filter_efficiency}) are placed in front of the sCMOS cameras (see Fig. \ref{fig:zyla_camera}). The filter exchanger mechanism allows  the desired filter to be placed in the optical path of the respective arm. Each filter exchanger has three available slots. The filters will be manually positioned during the first stage of operation, allowing the user to define the filters to be placed in each arm during an observation run according to Table~\ref{tab:filters}. For example, the filter exchanger in Arm 1 (FE1) has a mechanical scroll bar that can be moved to place either the filter $u'$, $g'$ or an 
``empty" slot in the optical path of Arm 1. The same mechanism has been implemented in the other two arms of OPTICAM. The aforementioned empty slots can also be used in future to house new filters for specific studies. As shown in Fig.~\ref{fig:optical_design}, the three optical paths of OPTICAM are optimised for their respective filter configurations. 

\begin{figure*}[!t]
	\centering
	\includegraphics[width=0.6\columnwidth]{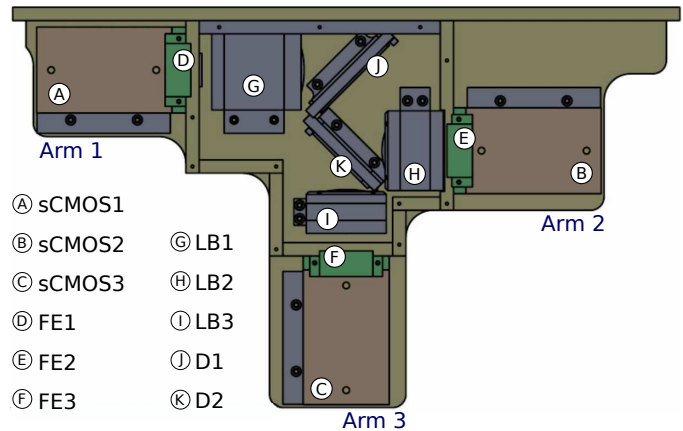}
	\caption{The mechanical structure of the OPTICAM instrument was designed according to the specific technical requirements of the OPTICAM project. Its three arms are optimised for $u'/g'$, $r'$ and $i'/z'$ filters respectively. The above figure indicates the presence of three {\zyla} sCMOS cameras (sCMOS1, sCMOS2 and sCMOS3, respectively; see Fig.~\ref{fig:zyla_camera}), the filter exchanger mechanism (FE1, FE2 and FE3), as well as the lenses within their respective barrels (LB1, LB2 and LB3) and dichroics (D1 and D2).}
	\label{fig:mechanical_config}
\end{figure*}

The current optical configuration will allow us to simultaneously observe the same FoV (see Table \ref{tab:opticam_fov}) with the 3 different sCMOS cameras, one in each arm, respectively.  The {\sc Zemax OpticStudio}\footnote{\url{https://www.zemax.com/products/opticstudio}} optical design software was used to simulate and accurately model the entire optical system of OPTICAM. An isometric projection of the system is shown in Fig. \ref{fig:mechanical_solidworks}, while a front-view is shown in Fig. \ref{fig:mechanical_config}. In both figures the three sCMOS cameras are shown in brown boxes, the filter exchange mechanism in green and the lenses and dichroics in grey. A thermal analysis was made every 5 {\textdegree}C from -10 {\textdegree}C to 20 {\textdegree}C. We found no significant variations in any of the optical cameras. As a consequence, the system is dominated by the thermal behaviour of the telescope. This thermal analysis revealed a good performance of the instrument, concluding that no thermal compensation is required within the OPTICAM operating limits.

The mechanical structure was designed using {\sc SolidWorks}\footnote{\url{https://www.solidworks.com/}} V2016. The design temperature and pressure was 20 {\textdegree}C and 1 atm, as the lenses would be manufactured and tested under these conditions. Static stress tests were carried out using the {\sc Autodesk Simulation Mechanical}\footnote{\url{https://www.autodesk.com/}} software. The main structure is being built in MIC-6 aluminium. The frame is made of 6061-T6 extruded aluminium alloy. DuPont Delrin\footnote{\url{http://www.dupont.com}} acetal homopolymer resin -- a highly-crystalline engineering thermoplastic -- has been used in the interfaces with the doublets of the lenses. Grade 316 stainless steel was used in some other minor components in order to guarantee a higher corrosion resistant structure.

\begin{table*}[!t]\centering
  \setlength{\tabnotewidth}{1.0\columnwidth}
  \tablecols{3}
  \setlength{\tabcolsep}{0.6\tabcolsep}
  \caption{Field-of-view of OPTICAM on the 2.1 m telescope} \label{tab:opticam_fov}
 \begin{tabular}{cccc}
    \toprule
    Arm & Filter & FoV (arcmin$^2$) & Plate Scale (\arcsec/pix)\\
    \midrule
    1 & ($u'$,$g'$) & 4.77$\times$4.77 & 0.1397\\
    2 & ($r'$)      & 4.80$\times$4.80 & 0.1406\\
    3 & ($i'$,$z'$) & 5.67$\times$5.67 & 0.1661\\
    \bottomrule
  \end{tabular}
\end{table*}

\section{Filters and dichroics} \label{sec:filter_efficiency}

\begin{figure*}[!t]
	\centering
	\includegraphics[width=0.5\columnwidth,angle=-90]{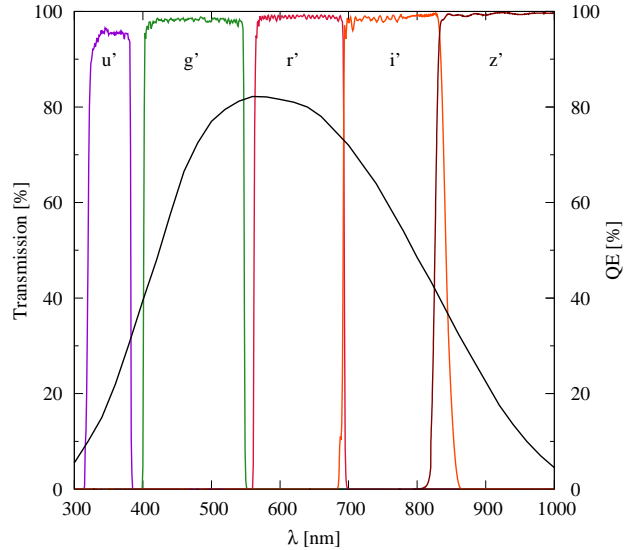}
	\caption{Transmission profiles of the OPTICAM SDSS filter-set (colour lines) and the QE curve (black solid line) of the {\zyla} sCMOS camera. OPTICAM will allow simultaneous imaging with 3 of the 5 SDSS filters (see Table \ref{tab:filters}).}
	\label{fig:filters_eficiency}
\end{figure*}

\begin{figure*}[!t]
	\centering
	\includegraphics[width=1.0\columnwidth,angle=0]{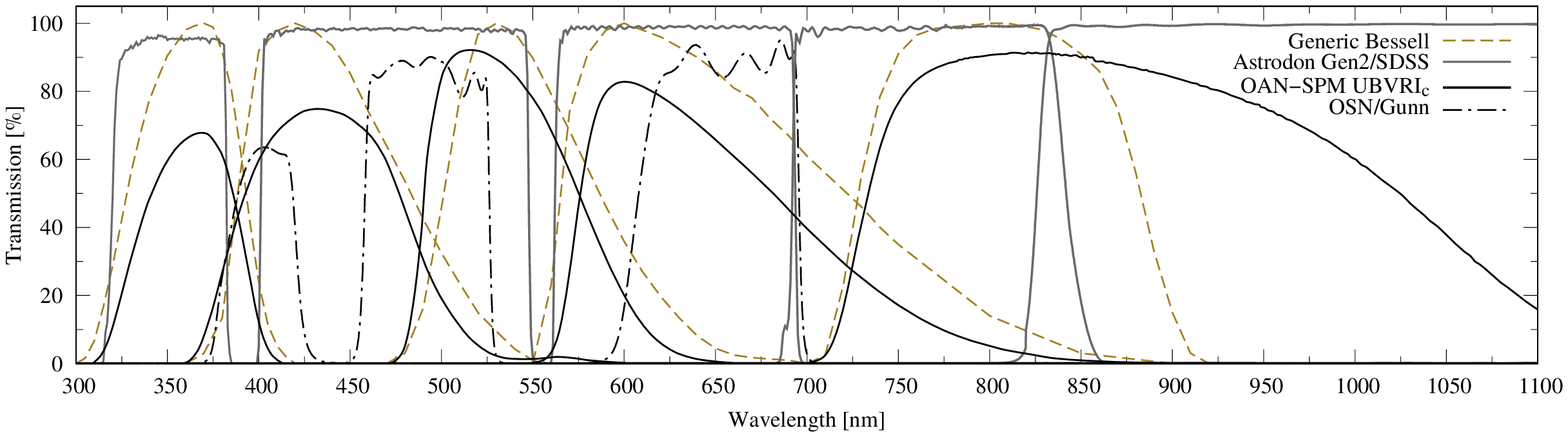}
	\caption{Filter curves of the SDSS filters compared with the Observatorio Sierra Nevada (OSN)/Johnson-Cousins ($U, B, V, R_{\rm C}$ and $I_{\rm C}$; black solid), Generic Bessell ($U, B, V, R, I$; dashed olive) and OSN/Gunn ($g$, $r$ and $i$; black dot-dashed) filter  transmission curves.}
	\label{fig:all_filters}
\end{figure*}

\begin{figure*}[!t]
	\centering
	\includegraphics[width=0.6\columnwidth,angle=270]{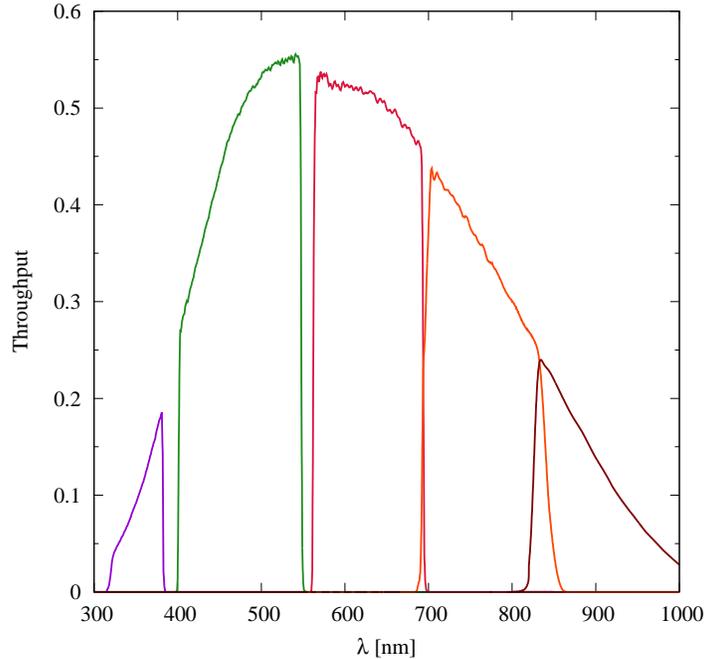}
	\caption{Throughput estimation of OPTICAM for the $u'$ (purple), $g'$ (green), $r'$ (red), $i'$ (orange) and $z'$ (black) bands. This calculation considers the contribution of filters, dichroics, camera lenses and the sCMOS detector.}
	\label{fig:total_transmission}
\end{figure*}

OPTICAM will have a set of SDSS filters \citep{fukugita+96}, which is the most common filter-set in modern Astronomy. These filters provide high-transmission, allowing for the detection of faint objects, and, at the same time, cover the entire optical wavelength range. We have acquired an Astrodon\footnote{\url{https://astrodon.com/}} Gen2 set of SDSS filters ($u'g'r'i'z'$) in order to fully cover the $320$--$1,100$ $\mathrm{nm}$ range. The transmission curves of the five broad-band filters and the quantum efficiency (QE) of the sCMOS camera are shown in Fig. \ref{fig:filters_eficiency}. This filter-set provides a peak transmission $>$ 95\%  ($>$90\% for $u'$). The filters $u'$, $g'$, $r'$, $i'$ and $z'$ passbands are shown in Table~\ref{tab:passbands}. A plot comparing the transmission curves of the Astrodon Gen2 SDSS filters with other widely-used filter-sets is shown in Fig.~\ref{fig:all_filters}. The total transmission curves for each photometric band of the OPTICAM system are shown in Fig.~\ref{fig:total_transmission}. Throughput estimations considered the combined action of the filters, the sCMOS detector and the dichroic beam splitters of the respective arms.

\begin{table*}[!t]\centering
  \setlength{\tabnotewidth}{1.0\columnwidth}
  \tablecols{3}
  \setlength{\tabcolsep}{0.8\tabcolsep}
  \caption{Astrodon Gen2 SDSS filter passbands} \label{tab:passbands}
 \begin{tabular}{ccc}
    \toprule
     Filter & Wavelength Range [$\mathrm{nm}$] & ${\Delta}{\lambda}[{\mathrm{nm}}]$\\
    \midrule
      $u'$    &  320--385 &  65\\
      $g'$    &  401--550 & 149\\
      $r'$    &  562--695 & 133\\
      $i'$    &  695--844 & 149\\
      $z'$    &  $>$820   & \nodata \\
    \bottomrule
  \end{tabular}
\end{table*}

OPTICAM will simultaneously use three different {\zyla} cameras, each one associated with its respective filter exchanger. The user will select the precise combination of filters to be used among the five available photometric filters at observation time according to Table \ref{tab:filters}. The addition of a $z_{s}$ filter, also from the SDSS filter-set, as well as a pair of narrow-band filters are being considered to allow secondary science projects to be carried out. Given the current instrumental configuration of OPTICAM, the use of narrow nebular filters (bandwidth $\sim$ 3 $\mathrm{nm}$), Galactic H${\alpha}$ and [OIII] is feasible for bright objects.

\begin{table*}[!t]\centering
  \setlength{\tabnotewidth}{0.5\columnwidth}
  \tablecols{4}
  \setlength{\tabcolsep}{0.5\tabcolsep}
  \caption{Filter options} \label{tab:filters}
 \begin{tabular}{cccc}
    \toprule
      \multicolumn{1}{c}{Filter Exchanger} & \multicolumn{3}{c}{Filter in each Position} \\
                      & Pos. 1 & Pos. 2 & Pos 3.\tabnotemark{a} \\
    \midrule
      FE1    &  $u'$  & $g'$    & \nodata   \\
      FE2    &  $r'$  & \nodata & \nodata   \\
      FE3    &  $i'$  & $z'$    & \nodata   \\
    \bottomrule
    \tabnotetext{a}{Empty positions can be used as void filter selections or, if required, future filter options will be included.}
  \end{tabular}
\end{table*}

The dichroic element is a fused silica substrate with optical interference coatings on both sides of the glass and a slight angle between its faces. Two dichroic beam splitters were designed for OPTICAM. The first dichroic (D1) has a size of $109~\times109~\pm0.2$ mm with a clear aperture of 96 mm in diameter, center thickness of 10 $\pm$0.1 mm and wedge angle $11\arcmin33\arcsec~\pm~30\arcsec$. The second dichroic (D2) has a size of $95~\times95~\pm0.2$ mm with a clear aperture of 86 mm in diameter, center thickness of 10 $\pm$0.1 mm and wedge angle $18\arcmin38\arcsec~\pm~30\arcsec$. Dichroics were designed to transmit and reflect an incoming beam at a 45\textdegree angle of incidence. The dichroics have a transition zone of 20 $\mathrm{nm}$ and a transmission of 97\% for $\lambda~>~$ 560 $\mathrm{nm}$ and 98\% for $\lambda~>~$ 700 $\mathrm{nm}$ respectively. Design requirements demanded a transmitted wavefront accuracy of $\lambda/4$ peak-to-valley across the whole clear apertures.

\section{{\zyla} cameras} \label{sec:camera}

\begin{figure*}[!t]
	\centering
	\includegraphics[width=0.3\columnwidth]{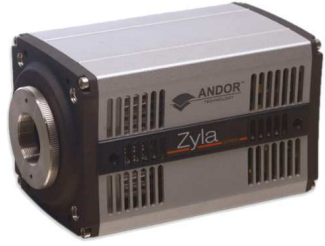}
	\caption{{\zyla} sCMOS camera. Image taken from the Andor Oxford Instruments website. }
	\label{fig:zyla_camera}
\end{figure*}

{\zyla} is a state-of-the-art sCMOS camera (see Fig. \ref{fig:zyla_camera}) sensitive in the 300--1,100 $\mathrm{nm}$ range, with a maximum QE of 82\% at 560 $\mathrm{nm}$ (see Fig. \ref{fig:filters_eficiency} for more details) and a linearity better than 99.8\% ($>$99.9\% for low-flux mode). This detector has a full-frame pixel array of 2,048$\times$2,048 pixels (6.5 $\mu$m pixels), a pixel well-depth of 30,000 $\mathrm{e^-}$ and exceptional data transfer efficiency. A more detailed list of specifications is shown in Table~\ref{tab:zyla_specifications}. The sCMOS camera is capable of delivering up to 53 frames per second (fps) in the 12-bit full-frame configuration and up to 40 fps in the 16-bit mode. By setting a smaller area of interest (AOI), also known as region of interest (ROI), it is possible to reach even higher readout speeds (see Table \ref{tab:frame_rates}). 

\begin{table*}[!t]\centering
\footnotesize
  \setlength{\tabnotewidth}{1.0\columnwidth}
  \tablecols{2}
  \setlength{\tabcolsep}{0.8\tabcolsep}
  \caption{{\zyla} camera model specifications\tabnotemark{a}.} \label{tab:zyla_specifications}
 \begin{tabular}{ll}
     \toprule
    \multicolumn{2}{c}{General specifications}\\
    \midrule
      Sensor type                         & Front Illuminated Scientific CMOS   \\
      Active pixels (W $\times$ H)               & 2,048 x 2,048 (4.2 Megapixel)   \\
      Pixel size                          & 6.5 $\mu$m   \\
      Sensor size                         & 13.3 x 13.3 mm  \\
      Pixel readout rate (MHz)            & Slow Read 216 (108 MHz x 2 sensor halves) \\
                                          & Fast Read 540 (270 MHz x 2 sensor halves)\\
      Read noise (e$^{-}$) median [rms]   & 0.90 [1.1] @ 216 MHz\\
										  & 1.10 [1.3] @ 540 MHz\\
      Maximum QE                          & 82 \%\\
      Sensor Operating Temperature        & 0 {\textdegree}C (up to 27 {\textdegree}C ambient)\\
      Dark current, e$^{-}$/pixel/sec @ min temp & 0.10 @ Air Cooled\\
      Readout modes                       & Rolling-Shutter\\
      Maximum dynamic range               & 33,000:1\\
      Pixel binning                       & Hardware binning: 2$\times$2, 3$\times$3, 4$\times$4, 8$\times$8 \\
      Data range                          & 12-bit \& 16-bit \\
      Interface options                   & USB 3.0 \\
      Hardware time-stamp accuracy        & FPGA generated timestamp with 25ns accuracy \\
      Linearity                           & $>$99.8 \% \\
    \bottomrule
    \tabnotetext{a}{More details: \url{https://andor.oxinst.com/products/scmos-camera-series/zyla-4-2-scmos}}
  \end{tabular}
\end{table*}

The {\zyla} camera has a due detector for applications requiring high sensitivity and high temporal resolution performance. In addition, this sCMOS camera has a thermoelectrically-cooled design, thereby avoiding the requirement of an active cooling system and high-cost maintenance, and a very low read-noise compared to CCDs. Due to its FPGA processing power and the memory capacity of the Andor sCMOS cameras, it permits implementation of bias offset compensation for every pixel in the array in real-time. This feature considerably lowers noise background into negligible levels, thus eradicating fixed pattern noise associated with CMOS cameras.

As shown in Table \ref{tab:frame_rates}, the maximum output frame-rate of the camera depends on both the size of the read pixel array and the bit-resolution. The pixel readout rate defines the rate at which pixels are read from the sensor. In the case of {\zyla}, a rolling-shutter\footnote{ \url{https://andor.oxinst.com/products/scmos-camera-series/zyla-4-2-scmos}} mode is used and the digitised signals are read out sequentially at a pixel readout speed of up to 540 MHz (270 MHz x 2 halves). In rolling-shutter mode -- contrary to the global-mode available in other camera models, where the whole pixel array is read out in a single scan -- different rows of the pixel array are exposed at different times as the rows are sequentially read from the center towards the upper and lower ends of each of the halves of the pixel array, respectively. OPTICAM uses as a default 540 MHz rolling-shutter configuration. In its fastest configuration at full-frame, the camera will be set up with rolling-shutter, overlap enabled and 12-bit mode. The sustained frame rate for this configuration is 53 fps with 100\% duty cycle due to the rolling-shutter capability (40 fps in 16-bit mode). This rolling-shutter mode allows a new exposure to start after each row of pixels has been read out, as a consequence, no time is lost between exposures.  

Higher frame rates can be achieved by reducing the desired pixel array size. Maximum sustainable frame rates and transfer rates for various (vertically centred) ROIs, at a 540 MHz readout speed, are also shown in Table \ref{tab:frame_rates}. The values shown in this table were obtained assuming a 1$\times$1 binning. In the fourth column of the same table, the byte-based transfer rate through the USB 3.0 port for the respective pixel array is shown. The {\zyla} cameras nominally operate in rolling-shutter mode. Since the actual write-speed of the host computer hard drive can limit the final sustainable frame rate, this effect has to be taken into account to avoid future data transfer malfunctions. In Table \ref{tab:opticam_fov}, detailed information is presented about the FoV and plate scales for OPTICAM, under different instrumental and AOI setups.

Image exposures can be started by a trigger event (e.g. TTL pulse or rising edge). We use a timing module described in \S~\ref{sec:control_data} to provide highly precise timing marks. The {\zyla} camera can be operated under room temperature conditions (up to 27 {\textdegree}C ambient), however, if its air-cooling system is software-enabled, the detector cools down, stabilising to a nominal operating temperature of 0 {\textdegree}C, regardless of the ambient temperature.

\begin{table*}[!t]\centering
  \setlength{\tabnotewidth}{0.65\columnwidth}
  \tablecols{5}
  \setlength{\tabcolsep}{0.50\tabcolsep}
  \caption{Maximum sustainable frame rates and transfer rates for various pixel arrays} \label{tab:frame_rates}
 \begin{tabular}{ccccc}
    \toprule
   Array Size\tabnotemark{a} & Frame Rate\tabnotemark{b} & BTR\tabnotemark{c} & Frame Rate\tabnotemark{b} & BTR\tabnotemark{c} \\
            & [fps@12-bit] & [Mb/s] & [fps@16-bit] & [Mb/s] \\
    \midrule
    2,048$\times$2,048   & 53   & 424   & 40   & 320    \\
    1,920$\times$1,080   & 107  & 423.2 & 80   & 316.4  \\
    512$\times$512     & 403  & 201.5 & 403  & 201.5  \\
    128$\times$128     & 1,578 & 49.3  & 1,578 & 49.3   \\
    \bottomrule
    \tabnotetext{a}{The pixel array (also called the area or region of interest) will be user-defined on the GUI window.}
    \tabnotetext{b}{Frame rates are given in fps (frames per second).}
    \tabnotetext{c}{Byte-based transfer rate (BTR).}
  \end{tabular}
\end{table*}

\section{Control and data-acquisition}\label{sec:control_data}
\begin{figure*}[!t]
	\centering
  \includegraphics[width=0.5\columnwidth]{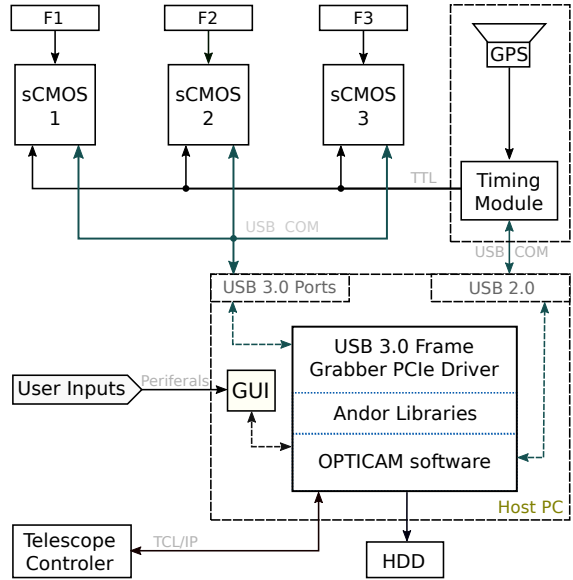}
  \caption{Block diagram of the OPTICAM control and data-acquisition system. Details in Sec. \ref{sec:control_data}.}
  \label{fig:block_diagram}
\end{figure*}

As shown in the block diagram of Fig. \ref{fig:block_diagram}, triple-band (filters F1, F2 and F3) simultaneous imaging will be possible using a set of three {\zyla} cameras (sCMOS1, sCMOS2 and sCMOS3, respectively). Image acquisition will be controlled from a single host computer. The software for image acquisition, flow control, storage and communication between interfaces and subprograms have been developed by collaborators of the present project, mainly written in C++. Specific software elements regarding frame-reading, image memory allocation, temperature control and other image-related handlers were adopted from the Andor {\sc Software Development Kit} (SDK\footnote{\url{https://andor.oxinst.com/products/software-development-kit/software-development-kit}}). The SDK provides us with a dynamic link library and a suite of functions entitled to configure the data-acquisition process from the Zyla sCMOS cameras used by OPTICAM.

Our instrumental control is closely related to the currently working OAN-SPM telescope control setups. The nominal communication standards between devices currently used at the OAN-SPM observatory have been adopted in order to deliver an instrument compatible with other systems now operating on the 2.1 m telescope. For OPTICAM, in the case of the graphical user interface (GUI), we have opted for a specific adaptation of the generic GUI \citep{colorado+14} currently used at the OAN-SPM, given our simultaneous imaging and frame rate needs. This new GUI will allow the observer to control the different frame-related parameters of the observation (e.g. exposure time, filters, ROI dimensions and binning) as well as the type of exposure to be acquired (e.g. test image, object, flat). Additionally, this interface will allow the quick visualisation of some important parameters of the exposed fields (e.g. object name, coordinates, airmass, observation date and time, camera details and filter data). 

High precision time stamps will be provided through a `Silverbox 2.0' timing module specifically designed on request for our project by the KTH Royal Institute of Technology\footnote{\url{https://www.kth.se/en}} staff members in Stockholm, Sweden. The module is intended for the generation of configurable and precisely synchronised timing pulses. Both the period of the pulse train and the pulse duration can be independently set to an integer factor of the base clock period. This module will be used by OPTICAM in order to acquire data from the 3 different cameras with a very accurate timing. The highly accurate clock of the Silverbox ($\pm$1 ppm precision; ppm stands for parts per million in relation to the crystal's oscillator nominal frequency) can be synchronised to either a
time-pulse from the Global Positioning System (GPS), or from a custom supplied, one-pulse-per-second line, or left free-running.

Assuming a selected full-frame (4.2 $\mathrm {Mpix}$), 12-bit resolution configuration (16-bit words for storage purposes; more details in \S~\ref{sec:camera}), each camera can provide up to 53 fps at maximum performance. Considering that each pixel is encoded with 16-bit words, this means that OPTICAM can deliver 8 ${\mathrm{Mb~s}}^{-1}$ through a USB 3.0 port. 

\section{Limiting-magnitude} \label{sec:limiting_magnitude}
\begin{figure}[!t]
	\centering
  \includegraphics[width=0.6\columnwidth]{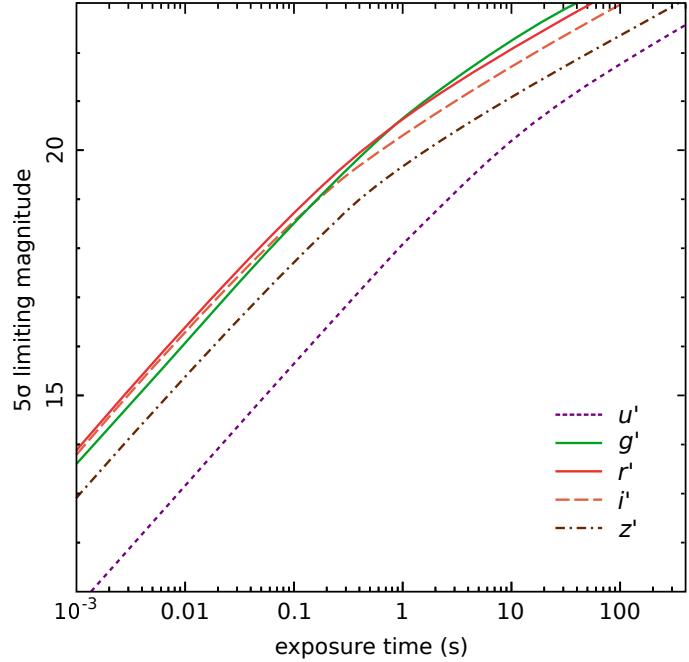}
  \caption{Preliminary limiting-magnitude plot for a detection by OPTICAM (at a signal-to-noise of 5-$\sigma$) in the OAN-SPM 2.1 m telescope.}
  \label{fig:limiting_magnitude}
\end{figure}

Fig.~\ref{fig:limiting_magnitude} shows a preliminary limiting-magnitude plot for a detection by OPTICAM (at a signal-to-noise of 5-$\sigma$) as a function of exposure time at the OAN-SPM 2.1 m telescope provided that the $f$/7.5 mount is installed. The curve was obtained assuming dark sky conditions\footnote{\url{http://www.astrossp.unam.mx/sitio/brillo_cielo.htm}} and an average seeing value of 0.7 $\arcsec$ (see \S~\ref{sec:spm_2m}). Atmospheric extinction was estimated from \citet{schuster+01}. Regarding the technical details of the Zyla cameras (e.g. QE and detector noise), we rely on the data provided by the {\zyla} user guide. The empirical throughput of OPTICAM at the 2.1 m telescope will be estimated during the first-light observation after some photometric standard stars are observed.

\section{Science with OPTICAM}\label{sec:science_opticam}

Due to its unique optical and timing capabilities, OPTICAM will allow us to perform strictly simultaneous, triple-band observations of a wide variety of astronomical objects with dynamical times-scales ranging from seconds in WDs to milliseconds in NSs and galactic stellar-mass BHs. The expected temporal resolution and accuracy achieved by OPTICAM will also allow the study of many other astrophysical objects including, amongst others, eclipsing binaries, exoplanets, active galactic nuclei, ultra-compact binaries, pulsars, pulsating WDs, stellar flaring events, and stellar occultations. In addition to stand-alone science, OPTICAM can complement simultaneous multi-wavelength campaigns of fast variability phenomena, or observations by several high-energy space observatories (e.g. {\it AstroSat}, {\it Chandra}, {\it NICER}, {\it NuSTAR}, {\it Swift} and {\it XMM-Newton}). Some of the proposed astrophysical objects and possible case-studies for use of this instrument are described below. 

\subsection{X-ray Binary Systems}
The study of BH and NS XRBs has rapidly evolved over the past two decades, largely due to the availability of powerful X-ray space observatories. The X-ray phenomenology of XRBs includes week-to-year-long outbursts, where their overall luminosity can increase by several orders of magnitude. However, most of the underlying physics is still under debate. An example of this is the presence of fast periodic/quasi-periodic variability (QPOs) observed at different wavelengths, often associated with orbital motions in the inner accretion disc and/or disc precession \citep[e.g.][]{ingram+16}. In this context, multi-wavelength observations of such fast-time variability might allow the degeneracy between models to be broken \citep{hynes+03,kalamkar+16,veledina+17}.

It is widely accepted that, given a continuous mass-transfer rate from the secondary star, outbursts of X-ray transients occur when a thermal instability occurs after the surface density at large radii reaches a critical value. From high-resolution, multi-wavelength observations of the binary BH V404 Cyg, \citet{kimura+16} found optical oscillations on time scales of 100 seconds to 2.5 hours, well-correlated with the X-ray emission. Since these oscillations occurred at mass-accretion rates much lower than expected, it was concluded that the mass-accretion rate is not the only key factor, particularly for sources with large discs, long orbital periods and large-amplitude oscillations \citep{kimura+16}. \citet{munoz-darias+16} reported the presence of rapidly-changing P-Cygni profiles due to winds, and attributed the X-ray and optical variability patterns to insufficient mass-flow reaching the innermost regions of the accretion disc. ULTRACAM simultaneous, three-band, sub-second observations of V404 Cyg showed a steep power density spectrum dominated by slow variations on $~$100--1000 s timescales \citep{gandhi+16}. These observations also showed persistent sub-second flares, particularly in the $r'$-band.
The origin of the observed lags between different wavelengths in accreting objects is also still unclear. Several promising physical processes have been put forward as explanations for such behaviour, e.g. reprocessing of the X-ray emission by the accretion disc, intrinsic variations of the thermal disc emission \citep{cameron+12} and changes in the projected direction to- and accretion flow into the compact jets \citep{bell+11}. At the median distances of $\sim$\,2\,kpc for BH XRBs \citep{gandhi+19}, the typical optical luminosities of the rapid sub-second flares alone can reach peak values of $\sim$\,10$^{36}$\,erg\,s$^{-1}$. Such powerful and rapid flares cannot originate as thermal emission. Thus, they appear to be important probes of non-thermal processes very close to the central compact objects.  
To explore some of these lines of research, OPTICAM will allow us to perform high-time resolution, simultaneous observations, made jointly with other high-performance space- and ground-based observatories, in order to provide a unique time-dependant, panchromatic view of XRBs. 

\subsection{Accreting White Dwarfs}
Interacting binary systems harbouring a WD are one of the most common accreting objects in the Galaxy. Due to their large number and relative proximity, accreting WDs are ideal laboratories to characterise the properties of accretion discs across a wide range of mass transfer rates. X-rays in cataclysmic variables (CVs) are thought to originate from the boundary layer \citep[see][for review]{kuulkers+06} or from the X-ray corona at low accretion rates \citep[e.g][]{king+84}. Roughly half of the gravitational potential energy is radiated from the aforementioned layer while the other half is radiated from the accretion disc through viscous processes \citep{dubrotka+14}.

Many CVs have short orbital periods of less than 4 $\mathrm{h}$ \citep[e.g.][]{patterson+02}. HS 0728+6738 is a CV with an orbital period (P) of 3.21 $\mathrm{h}$ showing optical variations on a time scale of $\sim$ 7 min. Additionally, the fast variability seen in HS 0728+6738 \citep{rodriguez+04} outside eclipse also shows QPOs on time scales of 10--30 minutes.High cadence observations of sources in eclipse with simultaneous time-resolved multi-band coverage, will allow us to dissect the different light components of these systems. Light curves obtained with OPTICAM will allow us to retrieve the geometry and orbital parameters of this type of system. Such techniques have already succeeded in detecting the long-sought population of WDs accreting from brown dwarfs \citep{littlefair+06,savoury+11,mcallister+17} which were later confirmed through spectroscopy \citep[e.g.][]{hernandez+16}.

\subsection{Active Galactic Nuclei}
For AGN harbouring a SMBHs in the range $10^{6}$--$10^{8}~{\rm M_{\odot}}$, the lags between various bands range from seconds to a few days \citep{breedt+10,cameron+12}. In particular, the study of low-mass AGNs, as in the case of Narrow-Line Seyfert 1s, requires a time-resolution on the order of a dozen seconds \citep[e.g.][]{mchardy+16}. An example is NGC 4051, a highly-variable AGN showing large variations in flux over relatively short timescales ($\sim$ hours) \citep[e.g.][]{mchardy+04,silva+16}. A major goal of such studies is the determination of inter-band lags, i.e. time-delays between different bands, which are thought to be due to the light-travel times between different parts of the system. The wavelength dependence of these lags can be used to infer the emissivity and temperature distributions across the accretion disc, as well as to estimate the size of the disc and its accretion rate. Thus, simultaneous coverage across multiple bands provided by OPTICAM will allow us to carry out this type of study.

\subsection{Eclipsing Binaries}
Eclipsing binaries are arguably the most important empirical calibrators in stellar astrophysics. In the case of short-period systems, their fundamental stellar and binary parameters (masses, radii, temperatures, luminosities and periods) can often be measured with both accuracy and precision. As an extreme example, high-precision photometric measurements of eclipse timings have even been used to establish the presence of an extra-solar planet orbiting the close binary NN Ser \citep{beuermann+10,marsh+14,parsons+14}. This makes short-period oscillation eclipsing binaries valuable subjects for both spectroscopic and multi-band photometric studies \citep[e.g.][]{lu+18}. The use of an instrument such as OPTICAM -- as part of simultaneous multi-wavelength coverage -- will make it possible to constrain with high-precision the physical parameters of a large number of eclipsing binaries.

\subsection{Exoplanets}
Although exoplanets were first studied using the radial-velocity technique, time-resolved colour photometry provides valuable astrophysical information that has led to detection and characterisation via transits of numerous exoplanets. Such measurements usually require extremely precise relative multi-band photometry and until recently they were only possible through the use of space telescopes or large ground-based facilities \citep{sedaghati+15}. However, precise {\it BVRIz'JH} observations by \citet{ricci+17} have recently confirmed the potential adequacy of small ground-based telescopes (in the range 36--152 cm) for extraplanetary transit research. A good example of this is the successful operation of MuSCAT-2 on the 1.52 Carlos S{\'a}nchez telescope and the current implementation of the TAOS-II\footnote{\url{https://taos2.asiaa.sinica.edu.tw/}} project at the OAN-SPM. Multi-colour, high-precision, fast-photometry obtained with OPTICAM will certainly help to increase the number of exoplanet observations at the OAN-SPM.

\section{Conclusions and Discussion}\label{sec:conclusions}

OPTICAM is a high-speed optical system designed to perform triple-channel fast-photometry. OPTICAM will have a set of $u'g'r'i'z'$ SDSS filters which will provide coverage in the $320<{\lambda}[\mathrm{nm}]<1,100$ wavelength range. OPTICAM will be mounted in the \add{Cassegrain focus} of the OAN-SPM 2.1 m telescope. Incident light will be split into three different beams using a pair of dichroic beam splitters. One beam is dedicated to either the $u'$ or $g'$ filter, whereas the second beam will be dedicated to $r'$ and the third beam to either the $i'$ or the $z'$ filter. These filter combinations, and additional empty positions, will be selected through the use of a manual filter exchanger available on each arm of the optical system. The filters will be placed according to the particular science case (within the limitations established in ~\S~\ref{sec:filter_efficiency}). The image acquisition will be carried out by three modern 2,048$\times$2,048 Andor Zyla sCMOS cameras, observing the same patch of  sky of approximately 5$\arcmin\times5\arcmin$ FoV (see Table \ref{tab:opticam_fov}).

The OPTICAM system will be capable of strictly simultaneous imaging, meaning that images will be acquired at exactly the same time through the use of a synchronisation card and dedicated software. OPTICAM will nominally allow sub-second imaging capabilities, nevertheless, higher readout speeds can be reached as a function of the ROI size. The three cameras will be synchronised with the aid of a precise timing module. Each image header will be time-stamped using this dedicated timing module equipped with a precise GPS system.

A theoretical limiting magnitude curve has been provided for the purposes of this paper. As far as we know, sky brightness and extinction values have not been measured before with the SDSS filter set at the OAN-SPM site. Transformations have been made from known $UBVRI_{\rm c}$ values, and extinction values were interpolated using the \citet{schuster+01} extinction curve as a first approximation. In order to estimate the empirical throughput of OPTICAM on the 2.1 m telescope and accurately characterise the response of the instrument, as well as to measure the average values of the sky brightness and extinction, a set of spectro-photometric standard stars need to be carefully observed at different air-masses. First on-site tests of the instrumental setup are expected to be carried out in October 2019. During this period, other, relatively well-known, bright objects will also be targeted, allowing us to make comparisons with published data of the same objects in order to establish performance comparisons with OPTICAM.

OPTICAM is a low-budget project compared to other, similar systems currently in operation (e.g. ULTRACAM and HiPERCAM). The sCMOS cameras had previously been acquired within said budgetary limitations. This pre-existing condition coupled with the scientific requirements established for OPTICAM, imposed very strict design requirements in order to provide a large FoV, simultaneous triple-band imaging and an appropriate plate scale for an instrument operating under the specific sky conditions of the OAN-SPM observatory. The design requirements are particularly oriented to get a good sampling of the point spread function of a point-like source without significant spacial degradation.

The most important limitation of OPTICAM is the low instrumental sensitivity in the bands $u'$ and $z'$. It will be in the observer's interest to decide when these filters should be used according to the needs of their underlying science, for example,  the acquisition times required (particularly for $u'$) to achieve a good S/N will limit the time-resolution possible (yet this is often required in some fast-photometry studies). Despite these limitations, in the context of conventional photometric studies, where timing capabilities are not critical, OPTICAM aims to be a very efficient instrumental system due mainly to its triple-band imaging capability, high time resolution and its well-defined detection band-passes. 

OPTICAM is a step forwards towards a new generation of instruments with high temporal resolution, which will enable studies of very fast astrophysical phenomena occurring in the range of milliseconds and seconds, a range that previously could not be achieved by conventional CCD photometric techniques. Due to its fundamental design, OPTICAM can be used to observe a variety of astrophysical sources such as XRBs, pulsating WDs, accreting compact objects, eclipsing binaries and pulsars. OPTICAM intends to match some of the capabilities of ULTRACAM, such as short exposure times, negligible dead-time and true simultaneous imaging with multi-wavelength coverage.

In order to maximise the long-term impact of the project, and its benefits to the astronomical community, we will ensure that all observations taken with OPTICAM become publicly available. Data will only be proprietary for six months after the observations are taken. In accordance with open-data policies, all observations will then be accessible to the entire international astronomical community. In addition, OPTICAM will serve as part of a world-wide transient network of instruments dedicated to the study of transient phenomena, such as SmartNET\footnote{\url{http://www.isdc.unige.ch/SmartNet/}} \citep{middleton+17}.

\bigskip

The authors would like to thank all the OPTICAM project members and collaborators. OPTICAM has been partially funded by the University of Southampton through the STFC Impact Acceleration Account grant and the Instituto de Astronom{\'i}a, UNAM. AC is supported by a UK Royal Society Newton International Fellowship. DA acknowledges support from the Royal Society. JVHS acknowledges support from a Vidi grant of the Netherlands Organization for Scientific Research (NWO) granted to N. Degenaar and a STFC grant ST/R000824/1. MM is supported by an Ernest Rutherford STFC fellowship. The SDSS set of filters was supplied by Astrodon. All lens polishing and barrels were provided by Tucson Optical. We thank the participation of N. Ivchenko and the KTH staff for the design of the timing module. We also thank V.~S. Dhillon, M. Richer and the Andor Support Team for their continuous technical support during the development of this project. We are thankful to the anonymous referee for constructive suggestions and comments.

\end{document}